\documentclass[preprint2]{aastex6}

\slugcomment{Accepted for publication in The Astrophysical Journal}
\usepackage{multirow}
\usepackage{amsmath}
\usepackage{booktabs}

\renewcommand{\vec}[1]{\boldsymbol{#1}}

\begin{document}


\title{The energy-dependent position of the \emph{IBEX} ribbon due to the solar wind structure}


\author{
Pawe\l{} Swaczyna,
Maciej Bzowski, 
Justyna M. Sok\'o\l{}
}
\email{pswaczyna@cbk.waw.pl}
\affil{Space Research Centre of the Polish Academy of Sciences (CBK PAN), Bartycka 18A, 00-716 Warsaw, Poland}
\turnoffedit

\begin{abstract}
Observations of energetic neutral atoms (ENAs) allow for remote studies of the plasma condition in the heliosphere and the neighboring local interstellar medium. The first \emph{IBEX} results revealed an arc-like enhancement of the ENA intensity in the sky, known as the ribbon. The ribbon was not expected from the heliospheric models prior to the \emph{IBEX} launch. One of the proposed explanations of the ribbon is the mechanism of the secondary ENA emission. The ribbon reveals energy-dependent structure in the relative intensity along its circumference and in the position. Namely, the ribbon geometric center varies systematically by about 10\degr{} in the energy range 0.7--4.3~keV. Here, we show by analytic modeling that this effect is a consequence of the helio-latitudinal structure of the solar wind reflected in the secondary ENAs. Along with a recently measured distance to the ribbon source just beyond the heliopause, our findings support the connection of the ribbon with the local interstellar magnetic field by the mechanism of the secondary ENA emission. However, the magnitude of the center shift in the highest \emph{IBEX} energy channel is much larger in the observations than expected from the modeling. This may be due to another, not currently recognized, process of ENA generation.  
\end{abstract}

\keywords{instrumentation: detectors -- ISM: atoms -- methods: data analysis -- solar wind -- Sun: heliosphere}



\section{Introduction}
\label{sec:introduction}

The Sun continuously emits an outward flow of plasma called the solar wind \citep{parker_1958}. The interaction of this flow with the partly ionized, magnetized local interstellar medium (LISM) creates a cavity in the interstellar matter called the heliosphere, with the heliopause as the boundary \citep{parker_1961}. The supersonic solar wind expands from the solar corona almost radially up to the termination shock, where its bulk speed decreases rapidly, and most of its kinetic energy is transfered into the internal energy of the plasma. The emerging plasma flow lines bend in front of the heliopause, which has an elongated, comet-like shape due to the relative motion of the Sun and the LISM.

The solar wind speed varies with heliographic latitude and time during the solar cycle. The solar wind can be investigated in-situ by spacecraft measurements or remotely by observations of interplanetary scintillations. During the solar minimum the slow wind occupies an equatorial band, while the fast wind is restricted to the polar caps. During the solar maximum the slow and fast streams of the solar wind are interspersed at all latitudes. The latitudinal structure of the solar wind with fast flow at high latitudes was inferred from the interplanetary scintillation observations \citep[e.g.,][]{kakinuma_1977,coles_1980} and observed in-situ by Ulysses \citep{phillips_1995,mccomas_2008}. 

Remote sensing of the plasma condition in the heliosphere and its neighborhood is carried out indirectly, by observations of energetic neutral atoms (ENAs) by \emph{Interstellar Boundary Explorer} \citep[\emph{IBEX},][]{mccomas_2009a}. Observations available from the \emph{IBEX}-Hi sensor \citep{funsten_2009} are carried out in several energy bands that cover a range 0.4--6 keV, i.e., energies typical for the solar wind.

The first sky maps obtained by \emph{IBEX} revealed an arc-like structure extending over a large part of the sky \citep{mccomas_2009,fuselier_2009,funsten_2009a}, dubbed the ribbon, not expected from simulations prior to the \emph{IBEX} launch \citep{schwadron_2009}. This discovery resulted in formulation of various hypotheses explaining the origin of the ribbon, which suggest the source region for the ribbon located at different regions of the heliosphere \citep{mccomas_2009,mccomas_2014a}. In a later analysis, \citet{swaczyna_2016} determined the heliocentric parallax of the ribbon, and thus the distance to the ribbon source at $140^{+84}_{-38}$~AU. This finding favors the hypothesis of the ribbon generation by the mechanism of the secondary ENA emission. In this mechanism, the primary ENAs, produced in the heliosphere, escape beyond the heliopause where, after two subsequent charge-exchange processes, they create a population of the secondary ENAs, from which a part is observed by \emph{IBEX} on the Earth's orbit \citep{mccomas_2009}. The highest signal is expected in the part of the sky where the lines of the local interstellar magnetic field, draped over the heliopause, are almost perpendicular to the lines of sight \citep{heerikhuisen_2010}.

The ribbon is observed in all \emph{IBEX}-Hi energy channels, but its intensity varies along its circumference and among the energy channels \citep{mccomas_2012a,funsten_2015}. \citet{funsten_2013} found that the positions of the maximum signal, obtained from the profiles across the ribbon, form shapes that may be approximated by circles or ellipses in the sky, and their centers in the sky systematically shift with energy. The intensity variation along the ribbon was qualitatively explained by \citet{mccomas_2012a} as due to the structure of the supersonic solar wind, which is the main contributor to the primary ENA flux, but the shift of the ribbon center by $\sim$10\degr{} between energies 0.7 keV and 4.3 keV remained unexplained. It was expected that the energy dependence of the charge-exchange reaction cross sections, and consequently of the distances to the secondary ENA source could explain this shift. However, \citet{zirnstein_2016} simulated this effect and showed that the shift in the ribbon center due to this effect should be $\sim$2\degr{} and occur only along the plane defined by the undisturbed magnetic field vector and the Sun velocity relative to the LISM, at odds with observation.
 
\edit1{In this paper, we use an analytical model of the secondary ENA mechanism supplemented with the helio-latitudinal structure of the solar wind. The fluxes calculated from this model are fitted} to follow a circle or an ellipse for each \emph{IBEX} energy channel (Section~\ref{sec:methods}). The fitted parameters are compared to the ones obtained in the data analysis by \citet{funsten_2013} (Section~\ref{sec:results}). The results strongly support the secondary ENA mechanism (Section~\ref{sec:conclusions}).

\section{Methods}
\label{sec:methods}

In the past, analyses of the secondary ENA emission were performed using both magnetohydrodynamic simulations \citep[e.g.,][]{heerikhuisen_2010}, and simplified analytics models \citep{mobius_2013,schwadron_2013,isenberg_2014}. Although details of these models are different, the main mechanism is the same. Namely, the primary ENAs produced in the heliosphere escape through the heliopause to the outer heliosheath where they are ionized, picked up by the draped interstellar magnetic field, and start to gyrate around the field lines. Eventually, they are re-neutralized via charge-exchange with ambient neutral atoms and some of them reenter the heliosphere. When the original ENA velocity is perpendicular to the magnetic field line, the guiding center of the created pick-up ion is pinned to the field line, and the resulting secondary ENA can be directed backwards to the Sun. These ENAs collectively form the ribbon. 

Here, we focus on the effect of the helio-latitudinal structure of the solar wind on the position of the ribbon. First, we model the flux of the primary ENAs originating from the supersonic solar wind (Section~\ref{sec:methods:nsw}), and subsequently this flux is used in the analytic model of the secondary ENA emission (Section~\ref{sec:methods:model}). \edit1{Based on the constructed model, we show the mechanism of the shift of the ribbon peak position (Section~\ref{sec:mechanismshift}).} The obtained signal is subsequently fitted to circles and ellipses (Section~\ref{sec:fitting}).

\subsection{Flux of the Neutral Solar Wind}
\label{sec:methods:nsw}
The primary ENAs are created both in the supersonic solar wind and in the inner heliosheath. \citet{zirnstein_2016} found that in magnetohydrodynamical models of the secondary ENA mechanism the contribution of the inner heliosheath ENAs can be neglected. Therefore, in this analysis we take into account only the contribution of the neutral solar wind (NSW) from the inner heliosphere.

The NSW is a supersonic solar wind, expanding inside the termination shock, that has been neutralized. The neutralization occurs mostly due to the charge exchange process between solar wind protons and the interstellar neutral atoms that have penetrated inside the termination shock. In this analysis, we take into account the helio-latitudinal structure of the supersonic solar wind. 

The observations of interplanetary scintillations collected by Institute for Space-Earth Environmental Research at Nagoya University \citep{tokumaru_2010} allow for determination of Carrington maps of the solar wind speeds. We use the solar wind structure as a function of heliographic latitude following a model by \citet{sokol_2012,sokol_2015c} based on the solar wind speed derived from the observations of interplanetary scintillations and in-situ in-ecliptic observations collected in the OMNI database \citep{king_2005}. This model provides the continuous in time and complete in latitude structure of the solar wind speed and density at 1 AU as a function of heliographic latitude and time from 1985 to 2013.

The supersonic solar wind is decelerated due to momentum loading into the plasma by ionization and charge exchange of the background neutrals. This solar wind slowdown was predicted theoretically \citep{fahr_2001,fahr_2002} and observed in situ by \emph{Voyager 2} \citep{richardson_2008}. We adopt a simple model of the inner heliosphere \citep{lee_2009}, in which the solar wind bulk speed $v$ is decreasing linearly with the distance to the Sun $r$:
\begin{equation}
 v(r)=v_0\left[1-\left(1-\frac{1}{2}\frac{\gamma-1}{2\gamma-1}\right)\frac{r}{\lambda_\mathrm{ml}} \right], \label{eq:decrvel}
\end{equation}
where $v_0$ is the solar wind bulk speed at 1 AU, $\gamma=5/3$ is the ratio of specific heats of the solar wind plasma, $r$ is the distance to the Sun, and $\lambda_\mathrm{ml}$ is the characteristic length for mass loading, given by the formulae:
\begin{align}
 \lambda_\mathrm{ml}&=\left(\lambda_\mathrm{cx}^{-1}+(n_0v_0)^{-1}(\nu_\mathrm{H} n_\mathrm{H}+4\nu_\mathrm{He} n_\mathrm{He})\right)^{-1},\\
 \lambda_\mathrm{cx}&=(\sigma_\mathrm{cx} n_\mathrm{H})^{-1}.
\end{align}
In these equations, $\sigma_\mathrm{cx}$ is the charge-exchange cross section between protons and hydrogen atoms \citep{lindsay_2005}, $n_\mathrm{H}$ and $n_\mathrm{He}$ are the number densities of the background interstellar neutral hydrogen and helium gas, $n_0$ is the solar wind density at 1~AU, and $\nu_\mathrm{H}$ and $\nu_\mathrm{He}$ are the photo-ionization rates for hydrogen and helium, respectively, at 1 AU \citep{bzowski_2013b,bzowski_2013c}. 

This model assumes that the background densities of neutral hydrogen ($n_\mathrm{H}=0.09\;\mathrm{cm}^{-3}$) \citep{bzowski_2008a} and helium ($n_\mathrm{He}=0.015\;\mathrm{cm}^{-3}$) \citep{gloeckler_2004} are constant in the inner heliosphere and equal to those at the termination shock. 
\edit2{In the reality, the neutral hydrogen density is not uniform and varies with the angle from the upwind direction and with distance from the Sun. The density of neutral hydrogen decreases for greater angles by a significant percentage \citep[e.g., see][]{heerikhuisen_2006, izmodenov_2013}. Additionally, the density is depleted inside $\sim$10~AU, and its structure is complex and evolves with time due to time-dependent ionization processes \citep{bzowski_2001, tarnopolski_2009}. In this analysis, we are interested in the NSW flux at the termination shock, regardless of the actual distances at which neutralizations occur. Therefore, important are the total column densities of the neutrals accumulated to the termination shock and the depletion for greater angles is partially compensated with the simultaneously increasing distance to the termination shock \citep{pogorelov_2009}. Moreover, the nonuniform distributions at a few AU from the Sun do not significantly affect the total column density. Consequently, the assumption of the constant densities is reasonable in the presented model.} 

\edit1{In this analysis, we build a time-averaged model, thus the probability distribution function of the NSW flux is constructed by averaging over a period of the solar activity. \citet{zirnstein_2015} found that the time delays between the primary ENA creation and observation of the secondary ENA range from $\sim$4 to $\sim$9 years, depending on the energy channel. Consequently, the secondary ENAs, observed by IBEX in years 2009--2011, i.e., the years used in the analysis by \citet{funsten_2013}, originate from the primary ENAs created between 2000 and 2007. Therefore, we averaged the NSW flux over solar cycle 23, which includes this interval.} The model of the solar wind we use has a time resolution of one Carrington Rotation (CR) \citep{sokol_2015c}, and in consequence, we average over the parameters obtained for the time range from CR 1909 to CR 2065. With this, we use the following formula for the NSW flux at the termination shock for heliographic latitude $\Theta$ and energy $E$:
\begin{align}
 I_\mathrm{NSW,TS}&(E,\Theta)=\frac{1}{N}\sum_{i=1}^{N}\int_{d_0}^{d_\mathrm{TS}} \frac{v_{0,i}(\Theta)n_{0,i}(\Theta)d_0^2}{d_\mathrm{TS}^2}\nonumber\\
 &\times\frac{e^{-r/\lambda_\mathrm{cx}}}{\lambda_\mathrm{cx}} \mathcal{N}\left(v_i(r),\delta v\big|\sqrt{2E/m}\right)\frac{1}{\sqrt{2mE}}dr. \label{eq:fluxnsw}
\end{align}
In this formula $i$ enumerates the parameters of the solar wind (density $n_{0,i}$ and bulk speed $v_{0,i}$ at 1~AU for requested latitude) for each selected CR ($N=156$)\footnote{The solar wind densities and speeds from \citet{sokol_2015c} are available as supplementary materials at \url{http://dx.doi.org/10.1007/s11207-015-0800-2}.}. Here we first generate the NSW flux for each CR, and then average the results, which is different than one would obtain by averaging the solar wind speed profile over the solar cycle first and calculating the NSW flux from the averaged solar wind later.

\begin{figure}
\epsscale{1.1}
 \plotone{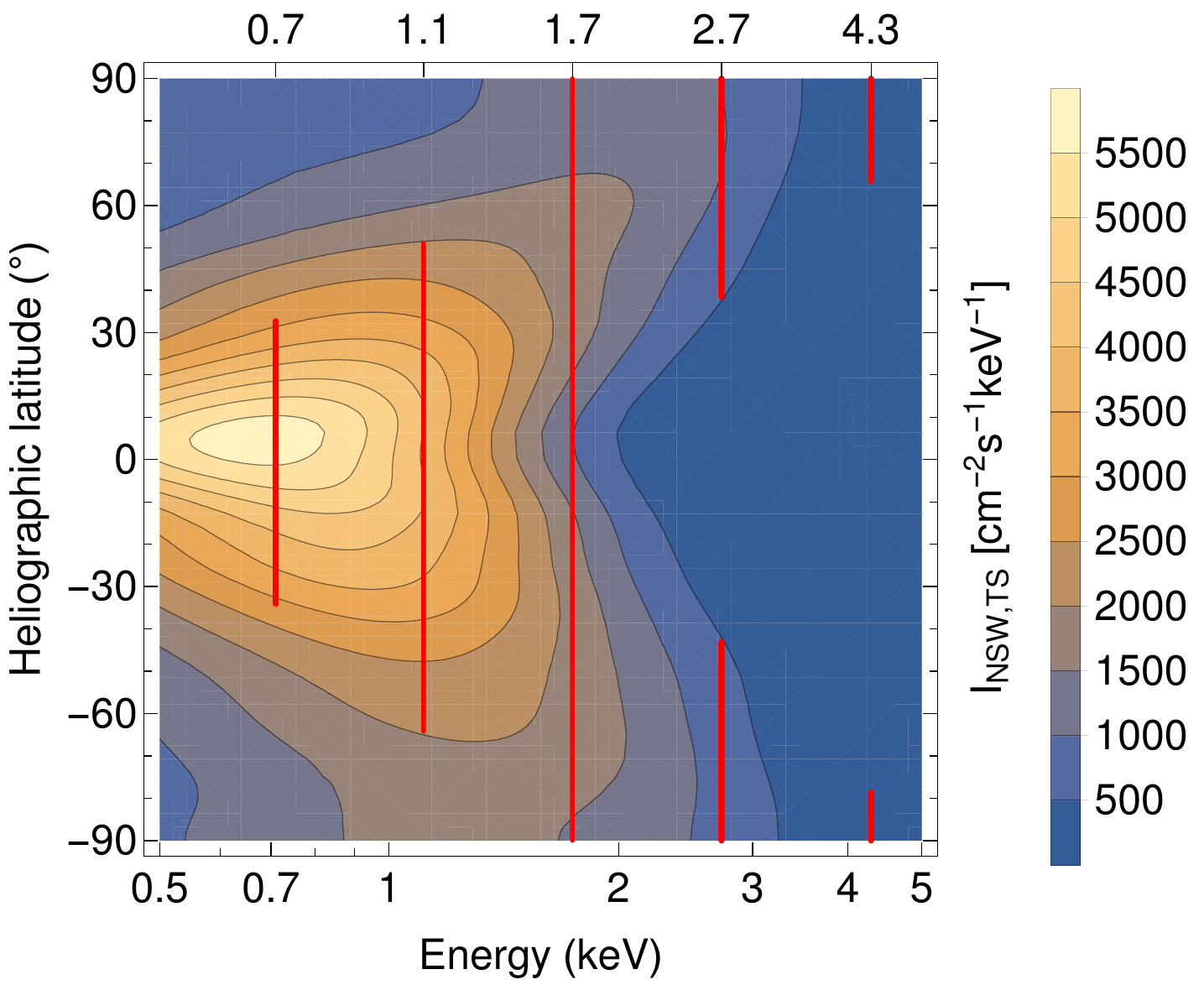}
 \caption{Differential NSW flux at the termination shock, obtained from averaging of the solar wind parameters for solar cycle 23. The characteristic energies are smaller at lower latitudes and larger at the higher ones. The red lines represent the heliographic latitude ranges for the centers of the \emph{IBEX}-Hi energy channels at 0.7, 1.1, 1.7, 2.7, and 4.3~keV, where the flux is larger than a half of the maximum for each channel.\label{fig:fluxnsw}}
\end{figure}

Independently of the charge exchange, the fluxes of the supersonic solar wind and the already created NSW decrease with distance, thus at the termination shock they need to be multiplied by the squared ratio of distances at 1~AU ($d_0$) and at the termination shock ($d_\mathrm{TS}$): $d_0^2/d_\mathrm{TS}^2$. The charge exchange process that is the source of the NSW is also responsible for the exponential decrease of the solar wind proton flux ($e^{-r/\lambda_\mathrm{cx}}$). The bulk speed $v_i(r)$ is decreasing according to Equation~\eqref{eq:decrvel}. We smooth the NSW speed distribution using the normal (Gaussian) distribution $\mathcal{N}$ with the mean value equal to the speed at the considered distance $v_i(r)$ and the standard deviation $\delta v=100$~km~s$^{-1}$ equal to the spread of the model speeds from \citet{sokol_2015c} and these observed by in-ecliptic spacecraft, collected in the OMNI database. We do so because in situ observations show that the velocity distribution function of the solar wind accumulated over the intervals of CR are much wider than it would be implied by the purely thermal spread of proton velocities. \edit1{The last term in Equation~\eqref{eq:fluxnsw} is the result of the conversion of variable. Namely, the normal distribution gives probability density in speed, thus we multiply it by $dv/dE=1/\sqrt{2mE}$ to get probability density in energy.} Figure~\ref{fig:fluxnsw} presents the differential flux NSW given by Equation~\eqref{eq:fluxnsw} as a function of energy and heliographic latitude. 

\subsection{Analytic model of secondary ENA emission}
\label{sec:methods:model}

In this analysis, we use an analytic model of the ribbon generation by the secondary ENA mechanism based on the observational constraints on the position and width of the ribbon in the sky. The model is an extension of the model by \citet{mobius_2013}. We employ the version of the model previously used in the assessment of the expected secondary helium ENA emission by \citet{swaczyna_2014}. With some rearrangement, the formula for the ENA differential intensity at \emph{IBEX} can be expressed as: 
\begin{equation}
 j_\mathrm{ENA}=\underbrace{\frac{1}{2\pi\Delta\psi}}_{G}\underbrace{\vphantom{\frac{1}{2\pi\Delta\psi}}\frac{n_\mathrm{NSW,TS}V_\mathrm{SW}}{\Delta E}}_{I_\mathrm{NSW,TS}} J(d_\mathrm{TS},d_\mathrm{HP},\lambda_\mathrm{H},\tilde{\lambda}_\mathrm{p}),\label{eq:jena}
\end{equation}
where
\begin{align}
 J(d_\mathrm{TS},d_\mathrm{HP},&\lambda_\mathrm{H},\tilde{\lambda}_\mathrm{p})=\frac{d_\mathrm{TS}^2}{\lambda_\mathrm{H}\tilde{\lambda}_\mathrm{p}} \int_{d_\mathrm{HP}}^{\infty} e^{-\frac{(r_1-d_\mathrm{HP})}{\lambda_\mathrm{H}}}\nonumber\\
 &\times\left[ \int_{r_1}^{\infty} e^{-\frac{(r_2-d_\mathrm{HP})}{\lambda_\mathrm{H}}}e^{-\frac{(r_2-r_1)}{\tilde{\lambda}_\mathrm{p}}}\frac{dr_2}{r_2^2}\right]dr_1. \label{eq:reflectance}
\end{align}
This form consists of three factors: the geometric factor $G$, the NSW flux at the termination shock $I_\mathrm{NSW,TS}$, and the dimensionless factor $J(d_\mathrm{TS},d_\mathrm{HP},\lambda_\mathrm{H},\tilde{\lambda}_\mathrm{p})$ that accounts for the ionization and re-neutralization of the NSW, with the inverse-square law for the NSW flux included. \edit1{The geometric factor had been originally expressed \citep{mobius_2013,swaczyna_2014} as $\Delta\psi/(2\pi\Delta\Omega)$, where $\Delta\Omega$ is the solid angle of the \emph{IBEX} field of view, and it is equivalent to the presented expression because $\Delta\Omega\approx\Delta\psi^2$.} Below, we describe the necessary modification of this formula in our analysis. 

In the original form it was assumed that the NSW flux is monoenergetic, so the flux was presented as a simple product of the density and speed of the NSW. The total flux was assumed to fit into a single \emph{IBEX} energy channel with the width $\Delta E$. Here, we replace this term with the differential NSW flux at the termination shock $I_\mathrm{NSW,TS}(E,\Theta)$, which depends on the energy $E$ and heliographic latitude $\Theta$. 

The factor $J( d_\mathrm{TS}, d_\mathrm{HP}, \lambda_\mathrm{H}, \tilde{\lambda}_\mathrm{p} )$ represents the effective part of the NSW that forms the secondary ENA. It is normalized to the flux of the solar wind at the termination shock. This is a convenient choice because the accumulation of the NSW ceases at the termination shock. In the formula $d_\mathrm{TS}$ and $d_\mathrm{HP}$ represent the distances to the termination shock and to the heliopause, respectively. The integrals run over $r_2$, which denotes the distance of the ionization of the primary ENA, and $r_1$, which denotes the distance where the re-neutralization occurs. The termination shock distance is assumed to be omni-directionally constant and equal to 90 AU, i.e., in the middle of the two distances of the termination shock crossing by \emph{Voyager 1} and \emph{Voyager 2} at 94 AU and 84 AU, respectively \citep{burlaga_2005,gurnett_2005,burlaga_2008,gurnett_2008}. For the heliopause distance we use a simple axisymmetrical model of the heliosphere with incompressible plasma flow by \citet{suess_1990}, for which we select the parameters so that the termination shock distance is 90~AU, and the distance to the heliopause at the \emph{Voyager 1} direction is 121~AU, as observed \citep{gurnett_2013}. With this model, the distance to the heliopause at the directions along the ribbon changes in the range 120--200~AU. The model used for the heliopause distance does not contain magnetic field and does not reconstruct the observed two-lobed structure of the heliotail \citep{mccomas_2013}. One of these lobes is coincident with the natural continuation of the ribbon location, and the ribbon signal is suppressed in this part of the sky. This effect is not reproduced by our model, but we drop this part of the ribbon from the analysis for the reasons described below. Finally, the heliopause distance is solely a function of the angular distance to the heliospheric nose, denoted as $\eta$, i.e., it is assumed to feature axial symmetry around the inflow direction.

\begin{figure}
\epsscale{1.1}
 \plotone{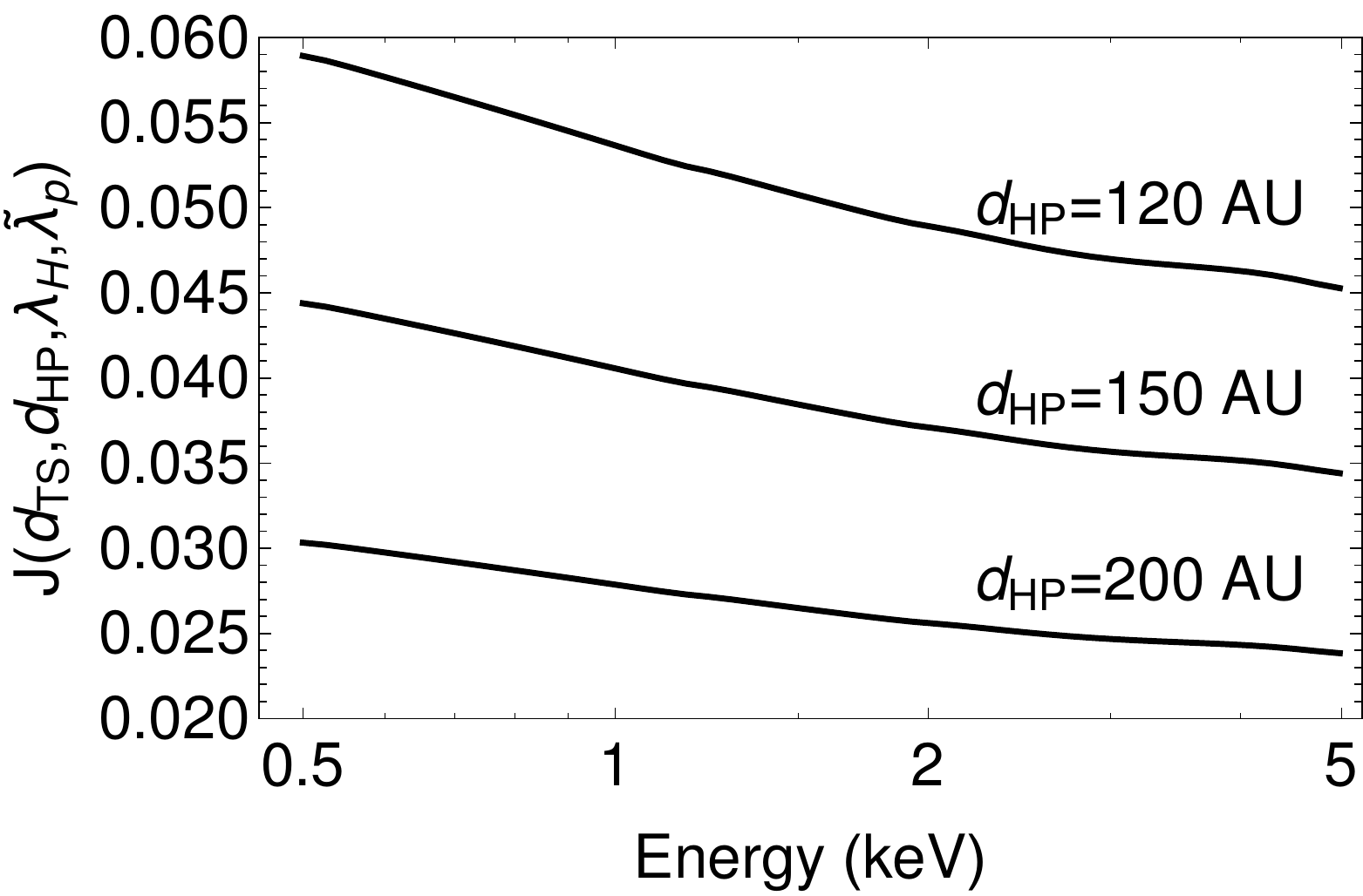}
 \caption{The factor $J( d_\mathrm{TS}, d_\mathrm{HP}, \lambda_\mathrm{H}, \tilde{\lambda}_\mathrm{p} )$ presented as a function of energy for three distances to the heliopause: 120 AU, 150 AU, and 200 AU. \label{fig:jfactor}}
\end{figure}

The mean free path for ionization of ENAs in the LISM $\lambda_\mathrm{H}$ and the effective mean free path for neutralization of the pick-up protons in the LISM $\tilde{\lambda}_\mathrm{p}$ depend on the considered energy and are given by the following formulae:
\begin{align}
 \lambda_\mathrm{H}(E)&=(\sigma_\mathrm{cx}(E)n_\mathrm{p}+\sigma_\mathrm{ion}(E)n_\mathrm{H})^{-1}, \label{mfpH}\\
 \tilde{\lambda}_\mathrm{p}(E)&=(\sigma_\mathrm{cx}(E)n_\mathrm{H})^{-1}\frac{V_\mathrm{Sun,LISM}|\sin\theta_{\vec{B},\vec{V}}|}{\sqrt{2E/m}},\label{mfpp}
 \end{align}
where $\sigma_\mathrm{cx}$ and $\sigma_\mathrm{ion}$ are cross sections for the charge exchange between hydrogen atom and protons \citep{lindsay_2005} and for the ionization of hydrogen atom by impact of another hydrogen atom \citep{barnett_1990}, respectively. The quantities $n_\mathrm{p}=0.06\;\mathrm{cm}^{-3}$ and $n_\mathrm{H}=0.2\;\mathrm{cm}^{-3}$ are the densities of protons and hydrogen atoms in the LISM \citep{frisch_2011}. The effective mean free path for protons also depends on \edit1{the velocity of the Sun in the LISM $V_\mathrm{Sun,LISM}=25.8\mathrm{\;km\;s}^{-1}$ \citep{bzowski_2015d} and the angle $\theta_{\vec{B},\vec{V}}$ formed by this velocity and the magnetic field direction. For this angle we adopt the value of 48\degr{} formed by the direction of the Sun motion from the analysis of interstellar neutrals \citep{bzowski_2015d} and energy-averaged center of the ribbon \citep{funsten_2013}.} The resulting value of the factor $J( d_\mathrm{TS}, d_\mathrm{HP}, \lambda_\mathrm{H}, \tilde{\lambda}_\mathrm{p} )$ is presented in Figure~\ref{fig:jfactor}. This factor has values in the range of 2.5--6\% with these assumptions, and moderately depends on the energy.

\edit1{The factor $J$ is a function of the physical properties of the outer heliosheath: the proton density, the hydrogen density, the velocity of the Sun, and the angle between this velocity and the magnetic field direction. We adopt the values for them as constant throughout the outer heliosheath. This is an approximation, but fortunately the factor $J$ is relatively robust. For example, the mean free paths given by Equations~\eqref{mfpH}~\&~\eqref{mfpp} with the presented values for energy 1.7 keV are $\lambda_\mathrm{H}(1.7\mathrm{\;keV})\approx780$~AU, and $\tilde{\lambda}_\mathrm{p}(1.7\mathrm{\;keV})\approx7.7$~AU. With the distance to the termination shock $d_\mathrm{TS}=90$~AU and to the heliopause $d_\mathrm{TS}=150$~AU, one obtains the value of $J=3.78\%$. Increasing or decreasing $\lambda_\mathrm{H}$ by 10\% result in the factor values of $3.54\%$ and $4.05\%$, respectively. For the same modification of $\tilde{\lambda}_\mathrm{p}$, the resulting values are $3.76\%$ and $3.81\%$. Consequently, the factor $J$ weakly depends on the physical conditions of the outer heliosphere. }

The geometric factor $G$ defines the solid angle into which the secondary ENAs are emitted. This factor effectively reflects the draping of the interstellar magnetic field and the creation and stability of the ring distribution of the pick-up primary ENAs that gyrate around the draped magnetic field line before re-neutralization. Using effective geometric factor is justified since it was found that the magnetic field draping alone cannot explain the energy dependence of the ribbon position \citep{zirnstein_2016}. Originally, the model assumed that the pick-up ions form a narrow cold ring distribution, such that the produced ENA fit entirely into the \emph{IBEX} field of view with the FWHM of $\Delta\psi=7\degr$ \citep{mobius_2013}. However, from the separation of the ribbon from the generally distributed flux it was found that the ribbon is broader \citep{schwadron_2011,schwadron_2014a}. Thus, we replace the formula presented in Equation~\eqref{eq:jena} with the following one:
\begin{equation}
 G(\phi)=\frac{1}{2\pi\sigma_\mathrm{rib}}e^{-\frac{(\phi-\phi_\mathrm{rib})^2}{2\sigma_\mathrm{rib}^2}}, \label{eq:geom}
\end{equation}
where $\sigma_\mathrm{rib}=10\fdg{}6$ is the standard width of the ribbon, which we adopt to reproduce the observed ribbon FWHM of 25\degr{} \citep{schwadron_2014a}. We assume that the geometric factor depends only on the distance $\phi$ to the ribbon center at ecliptic $(\lambda_\mathrm{rib},\beta_\mathrm{rib})=(219\fdg{}2,39\fdg{}9)$, with a maximum at the distance $\phi_\mathrm{rib}=74\fdg{}5$ \citep{funsten_2013}. We constructed this factor so that it does not depend on energy. Therefore, we adopt the parameters of the ribbon position obtained from averaging over energy channels. Consequently, the obtained shift of the fitted centers must be caused by the two others factors in Equation~\eqref{eq:jena}.

Finally, the intensity of the secondary ENA for energy $E$ in direction $\vec{\Omega}$ can be expressed as a product of three factors:
\begin{align}
 j_\mathrm{ENA}(E,\vec{\Omega})=&
 G(\phi)
 I_\mathrm{NSW, TS}(E,\Theta)\nonumber \\
 &\times J\left(d_\mathrm{TS},d_\mathrm{HP}(\eta),\lambda_\mathrm{H}(E),\tilde{\lambda}_\mathrm{p}(E)\right),
 \label{eq:intensity}
\end{align}
where $G(\phi)$ is the geometric factor given by Equation~\eqref{eq:geom}, $I_\mathrm{NSW, TS}(E,\Theta)$ is the NSW flux, given by Equation~\eqref{eq:fluxnsw}, and $J$ is the reflectance factor, defined as in Equation~\eqref{eq:reflectance}.

\begin{figure*}[t!]
 \plottwo{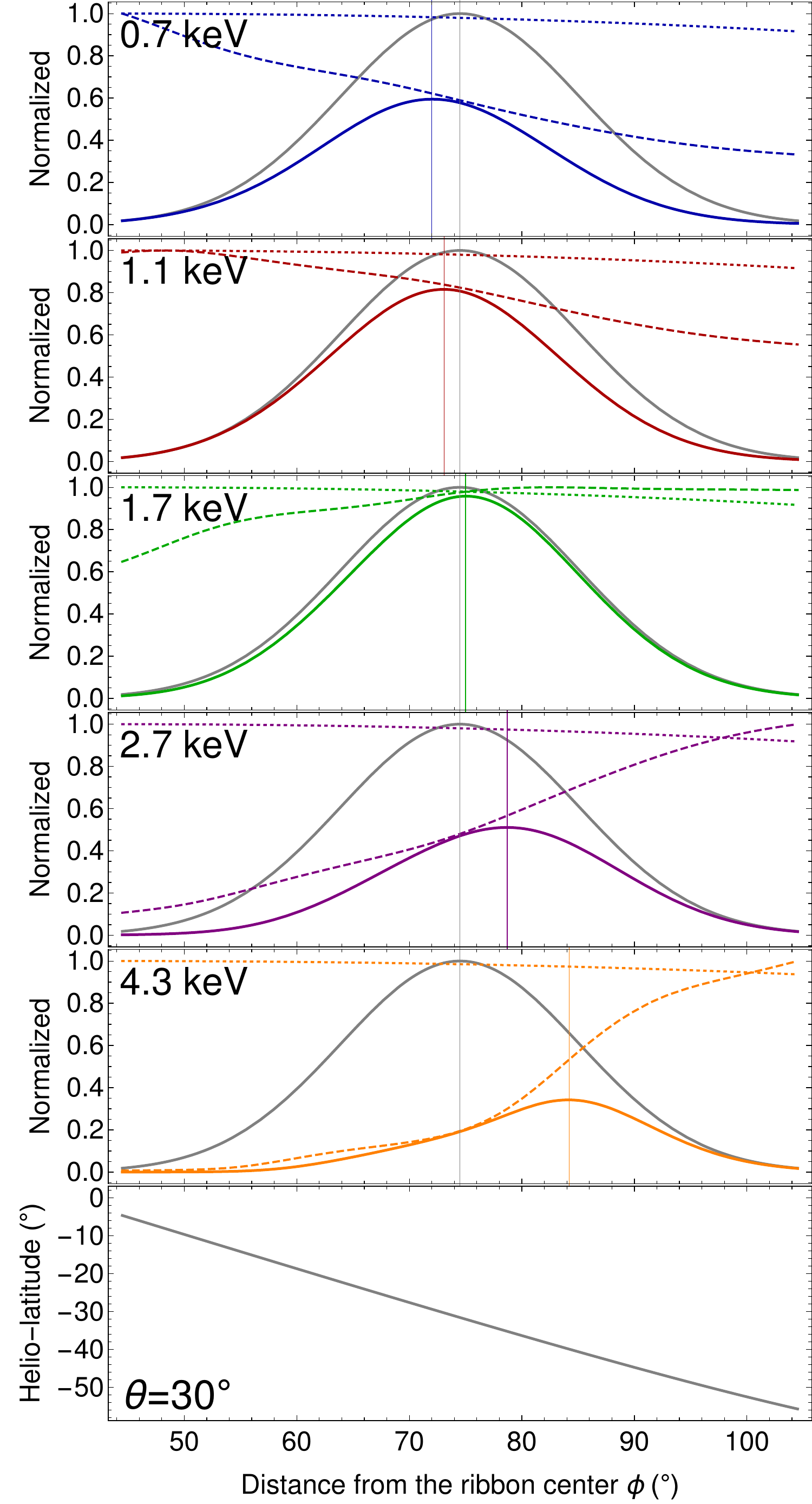}{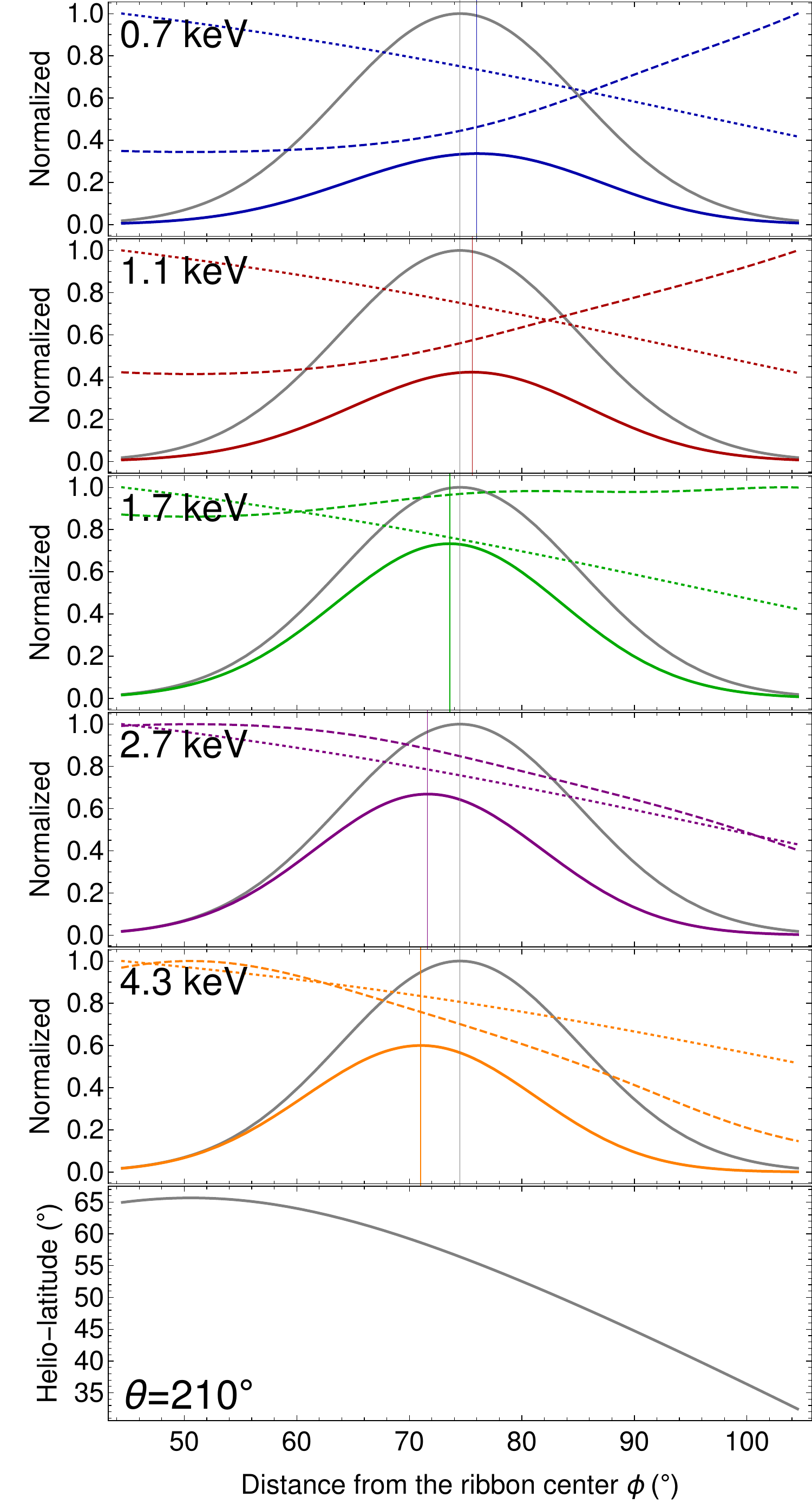}
 \caption{\edit1{Illustration of the influence of the helio-latitudinal structure of the NSW on the peak position of the secondary ENA emission responsible for the ribbon. Presented are the profiles in the ribbon coordinates for two exemplary azimuths (left panel: $\theta=30\degr$, right panel: $\theta=210\degr$) as a function of the distance to the ribbon center ($\phi$). Normalized values of the three factors forming the ribbon signal in the model (Equation~\eqref{eq:intensity}) are presented for the centers of the \emph{IBEX}-Hi energy channels: the geometric factor $G$ (gray line -- the same for each energy), the NSW flux  $I_\mathrm{NSW,TS}$ (dashed line), and the factor $J$ (dotted line). The products of these factors are proportional to the ribbon signal in the model and are presented in each panel with color solid line. The gradients of the heliographic latitudes (shown in the lowest panel) along the profile are reflected in the NSW fluxes, which cause the shift of the maximum positions (shown with vertical lines).\label{fig:profiles}}}
\end{figure*}

\subsection{Mechanism of the ribbon shift}
\label{sec:mechanismshift}

\edit1{Analysis of the ribbon peak position is the simplest in an auxiliary coordinate system with the pole close to the ribbon center. We follow \citet{funsten_2013} to construct such a coordinate system. Namely, we define a spherical coordinate system (hereafter the ribbon coordinates) so that the point $(\lambda_\mathrm{rib},\beta_\mathrm{rib})$ is the pole, and the prime meridian contains the direction of the heliospheric nose at $(\lambda_\mathrm{nose},\beta_\mathrm{nose})=(255\fdg{}8,5\fdg{}16)$ \citep{bzowski_2015d}. We denote the angular distance from the pole as $\phi$ and the azimuthal angle as $\theta$. \citet{funsten_2013} used a slightly different position of the pole and the meridian, but this does not influence the results of the presented analysis.}

\edit1{The mechanism of the ribbon shift can be tracked by the analysis of the relative contributions of the three factors forming the ribbon signal in Equation~\eqref{eq:intensity}. Figure~\ref{fig:profiles} illustrates how these factors vary along two exemplary azimuthal profiles for different energies. We normalize them so that their maxima at the presented range are equal to 1. The geometrical factor $G$ is the same for each energy by definition, and the variation of the normalized factor $J$ with energy is small. However, the NSW fluxes, which are functions of heliographic latitude, strongly influence the ribbon peak positions. The peak position is shifted in the same direction as the increase of the NSW flux. Consequently, there is systematic progression of the maxima of the secondary ENA intensity with increasing energy in each azimuthal profile. Effectively, they are combined to result in the progression of the ribbon centers with energy.}

\subsection{Fitting of circles and ellipses}
\label{sec:fitting}

We \edit1{calculate} the ENA intensity over the sky using the model of the secondary ENA emission presented above. Below, we describe the procedure used to find the circular and elliptic fits to the locations of the maximal signal along the ribbon. The procedure was tuned to follow the idea used previously by \citet{funsten_2013} to obtain the fits to the data. 

We integrate the signal given by Equation~\eqref{eq:intensity} over the $6\degr\times6\degr$ bins in the ribbon coordinates and over energies with the \emph{IBEX}-Hi energetic response function \citep{funsten_2009} for the respective energy channel. The same pixelization scheme was previously used in the data analysis by \citet{funsten_2013}. It is a different scheme from the scheme typically used to present \emph{IBEX} data, where the basis is the ecliptic coordinate system \citep{mccomas_2014b}. Subsequently, for each of the 60 meridian profiles we select 7 pixels so that the center pixel has the highest signal. The selected range of pixels is fitted to the Gaussian shape given by the formula: $A+B\exp(-(\phi-\phi_0)^2/(2\sigma^2))$. The fitted peak positions do not contain the uncertainties resulting from the statistical scatter, which is the main contributor to the total uncertainty in the analysis of the observations \citep{funsten_2013}. 

The fitted shapes (circles and ellipses) are not expected to reproduce the ribbon precisely. They are intended to be alternative, simplified descriptions of the ribbon morphology. In other words, we do not expect that with higher statistics, the location of the maximum signal of the ribbon will approach the position encircled by the fitted circle or ellipse. Consequently, we follow the selection of the pixels used in the original data analysis by \citet{funsten_2013}, and we need to weight the pixels to acknowledge the relative uncertainties of the fits to the data. 

Based on this prerequisite, we minimize the $\chi^2_\mathrm{C}$ and $\chi^2_\mathrm{E}$ estimators for the circular and elliptic models in the forms:
\begin{align}
 &\chi^2_\mathrm{C}(\vec{\Omega}_\mathrm{C},r_\mathrm{C})=\sum_{i}\frac{\left[g\left(\vec{\Omega}_i,\vec{\Omega}_\mathrm{C}\right)-r_\mathrm{C}\right]^2}{B_i^{-1}},\label{chi2c}\\
 &\chi^2_\mathrm{E}(\vec{\Omega}_\mathrm{E,1},\vec{\Omega}_\mathrm{E,2},a_\mathrm{E})\nonumber\\
 &=\sum_{i}\frac{\left[g\left(\vec{\Omega}_i,\vec{\Omega}_\mathrm{E,1}\right)+g\left(\vec{\Omega}_i,\vec{\Omega}_\mathrm{E,2}\right)-2a_\mathrm{E}\right]^2}{B_i^{-1}},\label{chi2e}
\end{align}
where $\vec{\Omega}$s represent the directions in the sky in whichever coordinate system, and $g$ returns the angular distance between the directions. The summation is over the ribbon positions $\vec{\Omega}_i$ in the azimuthal sectors enumerated by $i$, which run over the same sectors as those used in the data fitting by \citet{funsten_2013}. The quantities $B_i$ represent the heights of the ribbon profile, which we use for weighting. \edit1{This weighting is intended to recognize the relative uncertainties of the original data. Determinations of the peak positions in the data are a subject of uncertainties arising from the statistical scatter of the data. These uncertainties scale inversely proportional to the square root of the number of counts from the secondary ENA emission. The heights of the ribbon profile are proportional to number of counts if time of observations is uniformly distributed. We adopt this approach to stay in the fitting as close to the procedure adopted by \citet{funsten_2013} as possible.}

In the case of the circular fit, the parameters are the position of the ribbon center $\vec{\Omega}_\mathrm{C}$ and the ribbon radius $r_\mathrm{C}$. In the case of the elliptic fit, the parameters are the directions of the ellipse foci $\vec{\Omega}_\mathrm{E,1}$, $\vec{\Omega}_\mathrm{E,2}$ and the semi-major axis $a_\mathrm{E}$. Equivalently, the ellipse can be described by the following set of parameters: the center direction $\vec{\Omega}_\mathrm{E,0}$, the semi-major axis $a_\mathrm{E}$, the semi-minor axis $b_\mathrm{E}$, and the rotation angle $\theta_\mathrm{E}$. We also derive the eccentricity $e_\mathrm{E}$. We transform the fitted parameters to this set, since it was used for the data analysis by \citet{funsten_2013}. The rotation angle is given in the ribbon coordinates. 

\section{Results and discussion}
\label{sec:results}
\begin{figure*}
\plotone{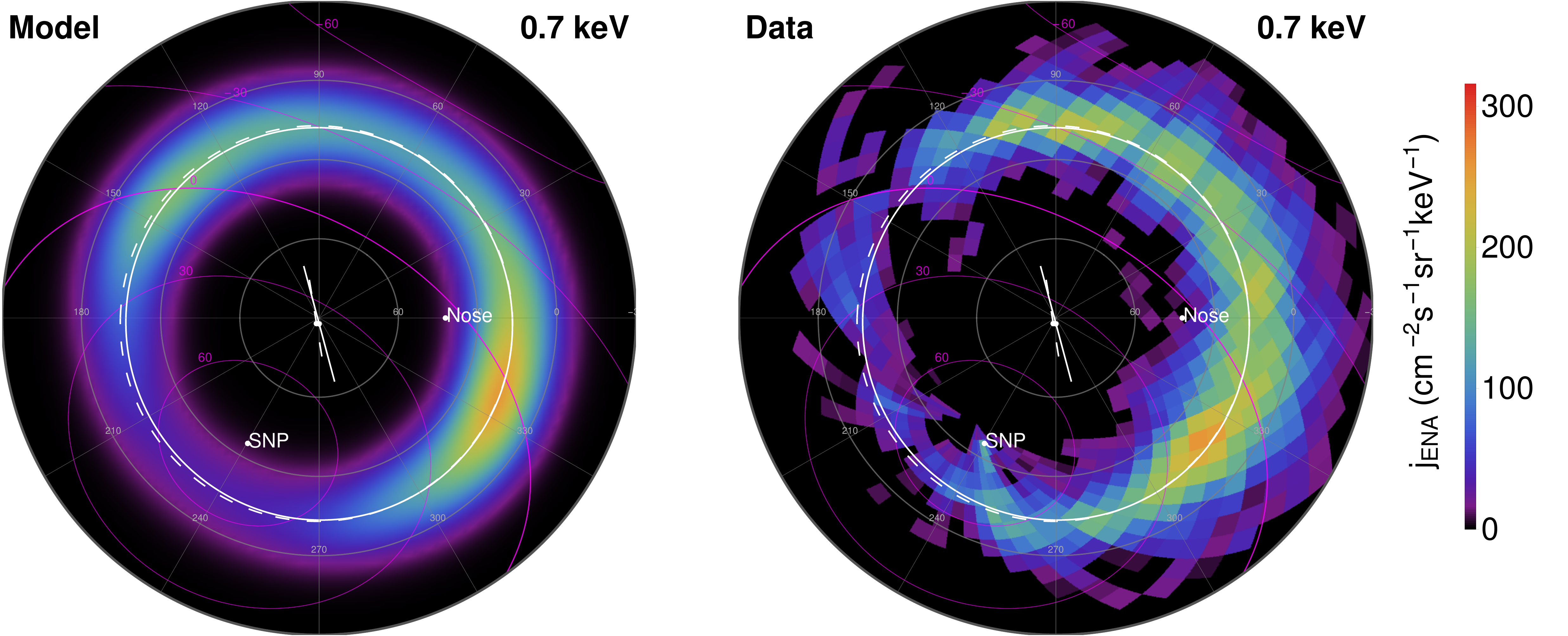}
\plotone{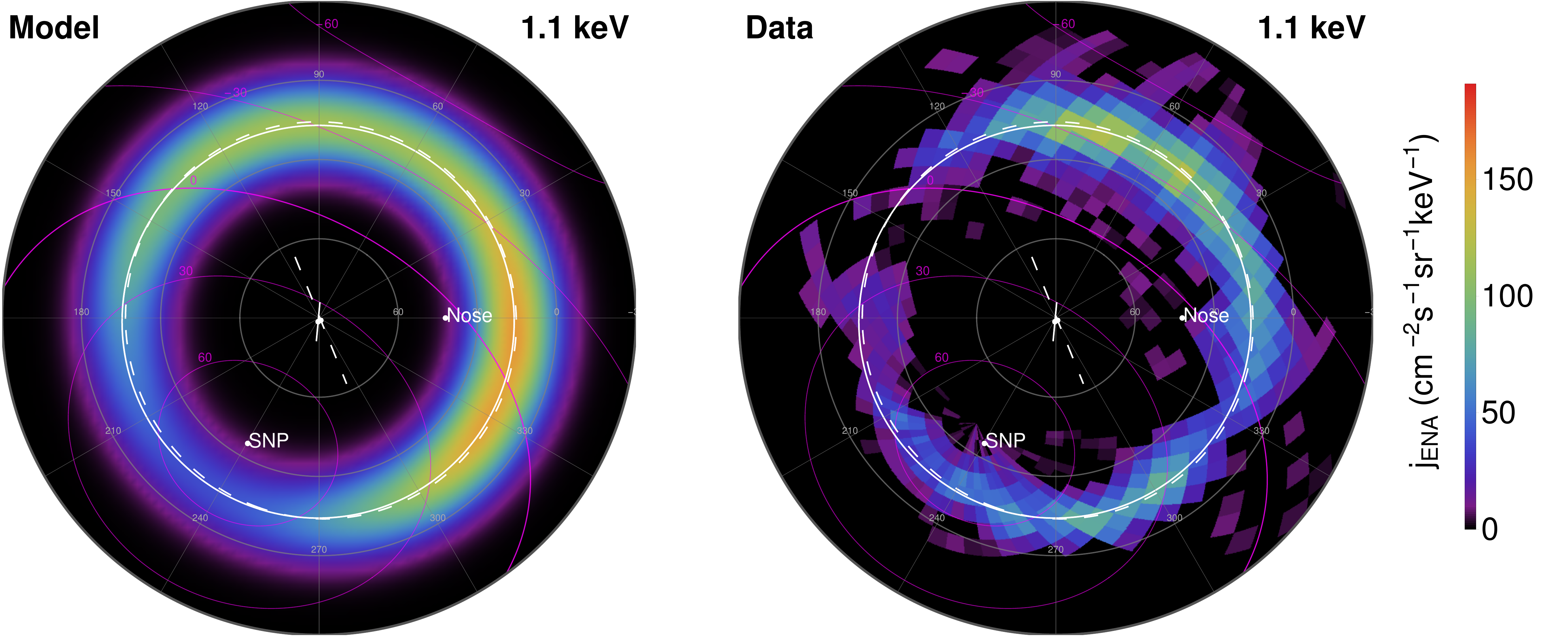}
\plotone{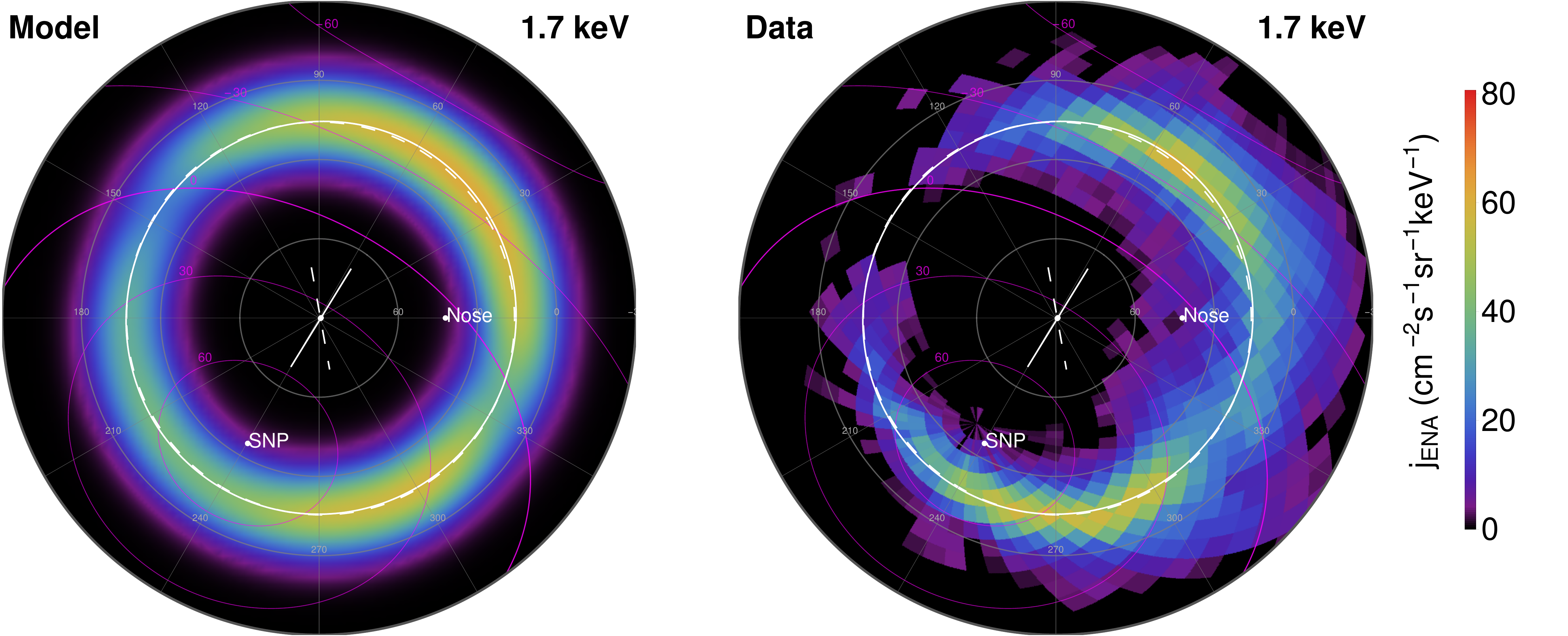}
 \caption{
 Maps of the \edit1{ribbon intensity calculated from the presented model} (left column) and the maps of the ribbon signal separated from the data \citep{schwadron_2014a} (right column) in the ribbon coordinates. The magenta lines mark the heliographic equator, and the $\pm$30\degr{} and $\pm$60\degr{} parallels. The white ellipses are the fits to the maximum signal along the ribbon for the \edit1{model} (solid line) and \edit1{reproduction of the fits to the data}  \citep[after][Table 2]{funsten_2013} (dashed line). The centers and the lines between the foci for both ellipses are shown with white points and lines, respectively. ``Nose'' marks the direction of the inflow of interstellar gas on the heliosphere \citep{bzowski_2015d}, and ``SNP'' the direction of the solar north pole. Top to bottom are the results for energy channels 0.7, 1.1, and 1.7 keV.
 \edit1{The color scheme for each energy channel is shown on the right and is common for the model and data.}
 \label{fig:maps234}}
\end{figure*}

\begin{figure*}
\plotone{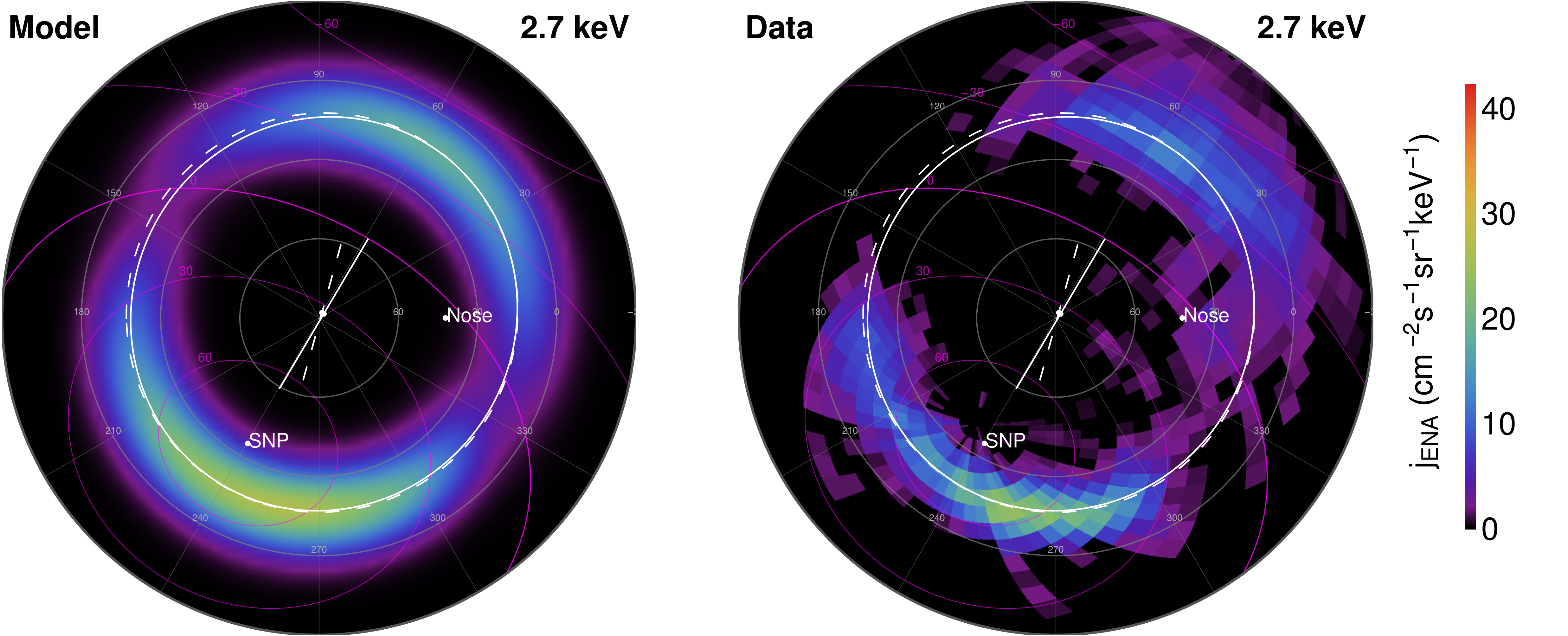}
\plotone{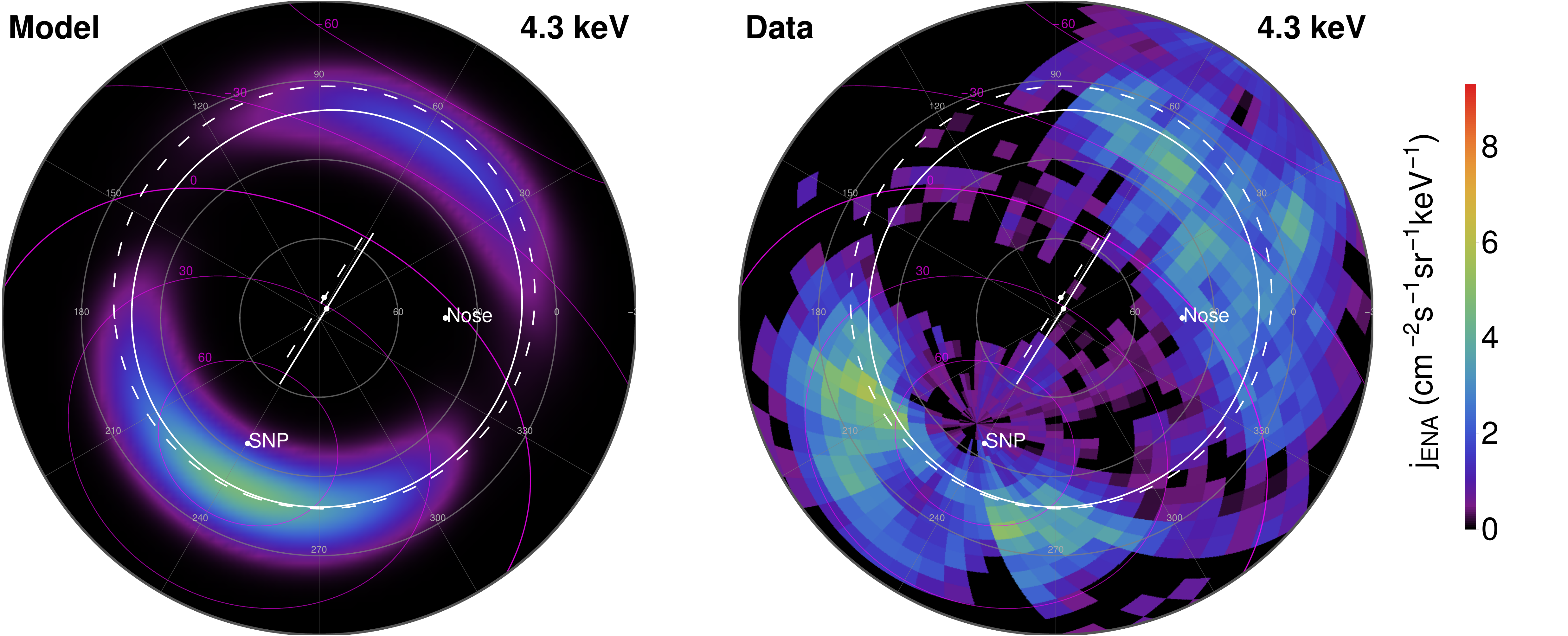}
 \caption{As in Figure~\ref{fig:maps234}, but for energy channels 2.7 and 4.3 keV. \label{fig:maps56}}
\end{figure*}

The \edit1{signal calculated from the presented model} is compared with the signal extracted from the observations by \citet{schwadron_2014a}\footnote{The numerical values of ribbon signal extracted from the data are available as \emph{IBEX} Data Release 8 \cite{schwadron_2014a} at \url{http://ibex.swri.edu/ibexpublicdata/Data_Release_8}.} in Figures~\ref{fig:maps234}~\&~\ref{fig:maps56}. The maps are plotted in the ribbon coordinates, so they can be compared also with the maps in the previous analysis by \citet[][Figures 2 \& 3]{funsten_2013}. In Table~\ref{table:fitcompare} we compare these parameters obtained from our \edit1{model} fitting and the one found by \citet{funsten_2013} from data analysis. The mean deviations between the ribbon locations and the fitted ellipses and circles are denoted as $\sigma_\mathrm{E}$ and $\sigma_\mathrm{C}$, respectively.

\begin{deluxetable*}{ccrrrrrrrrrrr}
 \tablewidth{0pt}
 \tablecolumns{13}
 \tablecaption{Comparison of fitted parameters \label{table:fitcompare}}
 \tablehead{
    \colhead{}		&
    \colhead{}			&
    \multicolumn{7}{c}{Elliptic fit}	&
    \multicolumn{4}{c}{Circular fit}	\\
    \cmidrule(lr{.75em}){3-9} \cmidrule(lr{.75em}){10-13} 
    \multicolumn{2}{c}{$E$ (keV)}		&
    \colhead{$\lambda_\mathrm{E}$~(\degr)}			&
    \colhead{$\beta_\mathrm{E}$~(\degr)}		&
    \colhead{$\theta_\mathrm{E}$~(\degr)} &
    \colhead{$a_\mathrm{E}$~(\degr)} &
    \colhead{$b_\mathrm{E}$~(\degr)} &
    \colhead{$e_\mathrm{E}$}				&
    \colhead{$\sigma_\mathrm{E}$~(\degr)}	&
    \colhead{$\lambda_\mathrm{C}$~(\degr)}	&
    \colhead{$\beta_\mathrm{C}$~(\degr)} &
    \colhead{$r_\mathrm{C}$~(\degr)} &
    \colhead{$\sigma_\mathrm{C}$~(\degr)}
 }
 \startdata
 \multirow{2}{*}{0.7} 	& d\tablenotemark{a} & 219.8 & 42.2 & 97.4 	& 74.9 & 73.2 & 0.22 & 1.4 & 218.5 & 43.1 & 74.8 & 2.1 \\
						& m 	& 221.0 & 41.7 & 103.0 	& 74.4 & 71.0 & 0.30 & 0.3 & 220.2 & 42.5 & 74.3 & 0.3 \\[1.1ex]
 \multirow{2}{*}{1.1} 	& d 	& 220.6 & 40.2 & 111.3 	& 75.4 & 71.0 & 0.34 & 1.8 & 220.3 & 40.5 & 73.3 & 2.4 \\
						& m 	& 219.9 & 41.3 & 83.5	& 74.4 & 74.0 & 0.09 & 0.3 & 219.9 & 41.3 & 74.3 & 0.3 \\[1.1ex]
 \multirow{2}{*}{1.7} 	& d 	& 219.9 & 39.7 & 100.0 	& 74.4 & 71.8 & 0.26 & 1.5 & 219.6 & 39.8 & 73.2 & 1.7 \\
						& m 	& 219.9 & 39.4 & 58.5 	& 74.7 & 71.5 & 0.29 & 0.6 & 219.4 & 39.4 & 74.4 & 0.7 \\[1.1ex]
 \multirow{2}{*}{2.7} 	& d 	& 218.8 & 37.6 & 76.3 	& 75.7 & 70.9 & 0.35 & 1.8 & 217.9 & 37.7 & 74.4 & 2.2 \\
						& m 	& 219.9 & 37.6 & 60.8 	& 75.3 & 67.8 & 0.43 & 0.5 & 218.9 & 37.6 & 74.8 & 0.8 \\[1.1ex]
 \multirow{2}{*}{4.3} 	& d 	& 215.5 & 32.5 & 65.3 	& 80.3 & 75.7 & 0.33 & 2.9 & 214.2 & 32.4 & 79.2 & 3.0 \\
						& m 	& 219.5 & 35.4 & 61.6 	& 75.9 & 68.2 & 0.44 & 0.8 & 218.6 & 35.4 & 75.5 & 1.0 \\
  \enddata
  \tablenotetext{a}{`d' denotes the parameters from the fitting to the data obtained by \citet{funsten_2013}, and \edit1{`m' to the signal calculated from the presented model} (this analysis).} 
\end{deluxetable*}

\begin{figure*}
\plottwo{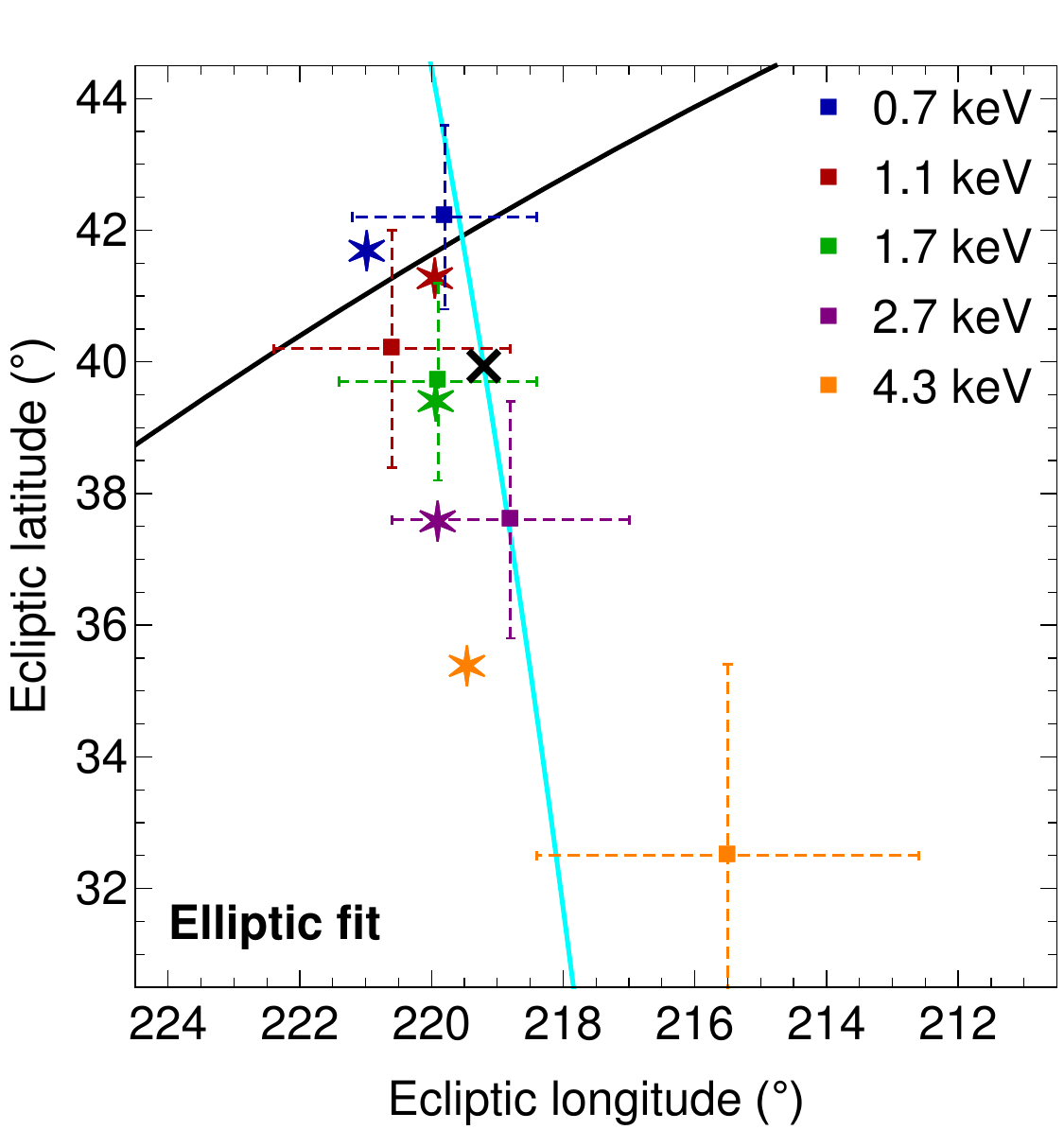}{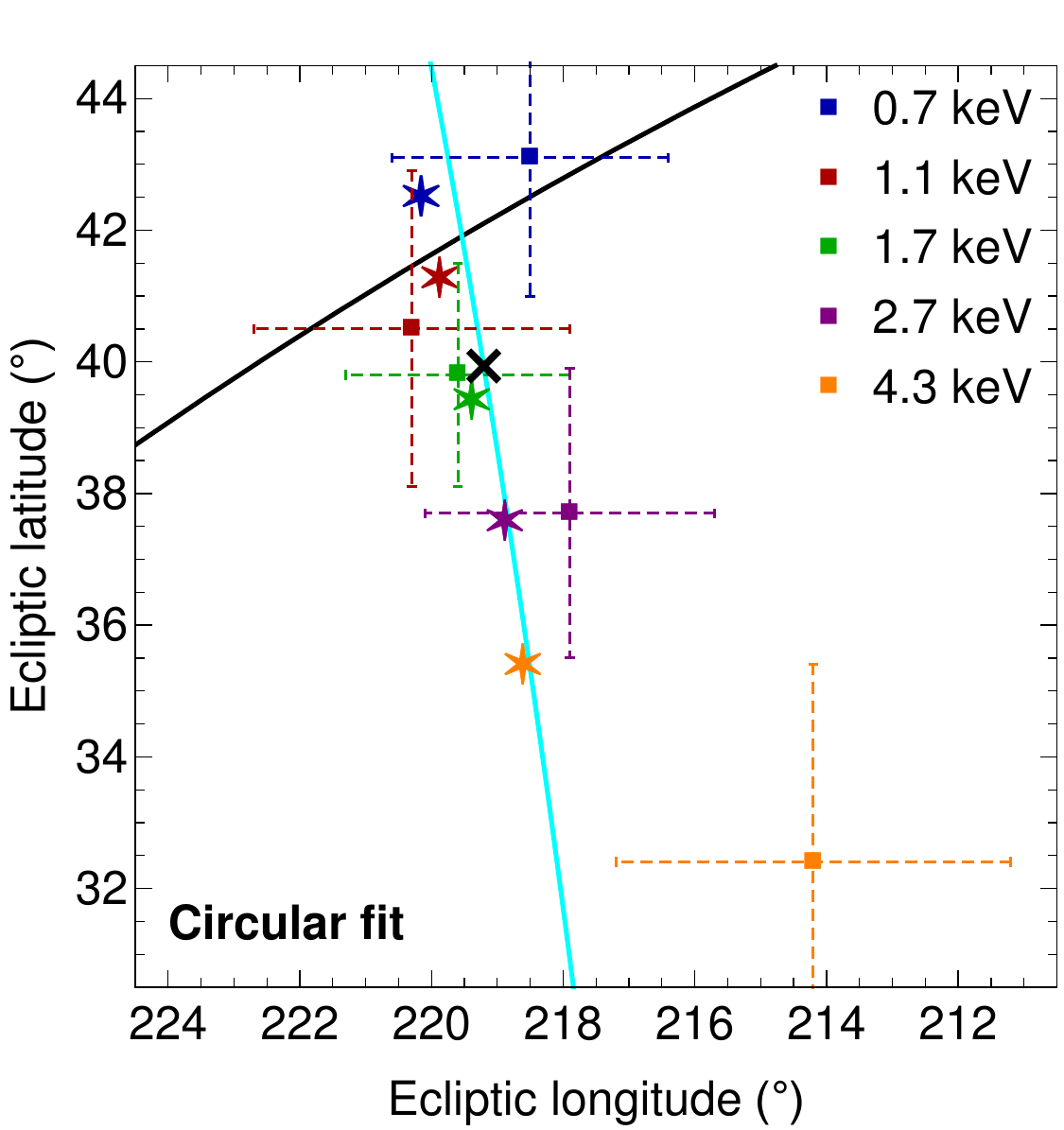}
 \caption{Centers of the ellipses (left panel) and circles (right panel) fitted to the data \edit1{by \citet{funsten_2013}} (symbols with error bars) and to the \edit1{signal calculated from the model in this analysis} (stars) for different \emph{IBEX}-Hi energy channels. The error bars for the data fitting are adopted as in the original analysis. The black line represents the interstellar neutral deflection plane \citep{kubiak_2016}. The \edit1{cyan} line is the heliographic meridian that contains \edit1{the solar poles and} the energy-averaged center of the ribbon $(\lambda_\mathrm{rib},\beta_\mathrm{rib})$, marked as $\times$. \label{fig:fitcenters}}
\end{figure*}

We compare the fitted centers in Figure~\ref{fig:fitcenters}. The displacements of the fit centers between subsequent energy channels obtained from the \edit1{model} match those obtained by \citet{funsten_2013} from the data analysis. The energy sequence is not aligned along the plane that includes the interstellar neutral flows and the ribbon center \citep{kubiak_2016}, known as the neutral deflection plane (black line in Figure~\ref{fig:fitcenters}), but it is approximately parallel to \edit1{the great circle intersecting the solar poles and the energy-averaged center of the ribbon, i.e., the local heliographic meridian (cyan line). The uncertainty analysis done by \citet{funsten_2013} was simplified, so the $\sigma_\mathrm{E}$ parameters were adopted as the uncertainties of the ribbon centers.} Such a procedure probably overestimated the actual uncertainties. Consequently, we are not able to formally check the consistency between the data and \edit1{model} results. The ellipticities expressed by the rotation angle and eccentricities are similar for the observations and \edit1{model}, even though we have assumed a simple circular shape for the geometric factor. 

The circular fits are intended as a sanity test for our baseline results, obtained from the elliptic fits: a qualitative difference between the two models would cast doubt on the credibility of our conclusions. But the results from the circular fits are similar to the elliptic fits. In the case of circular fits, the centers are even better aligned with the local heliographic meridian. Comparing the mean differences of the peak location to the fitted signal ($\sigma_\mathrm{E}$ vs $\sigma_\mathrm{C}$), it is visible that elliptic fits are better. This is a natural consequence, since the ellipse is generalization of the circle. 

\begin{figure}
\plotone{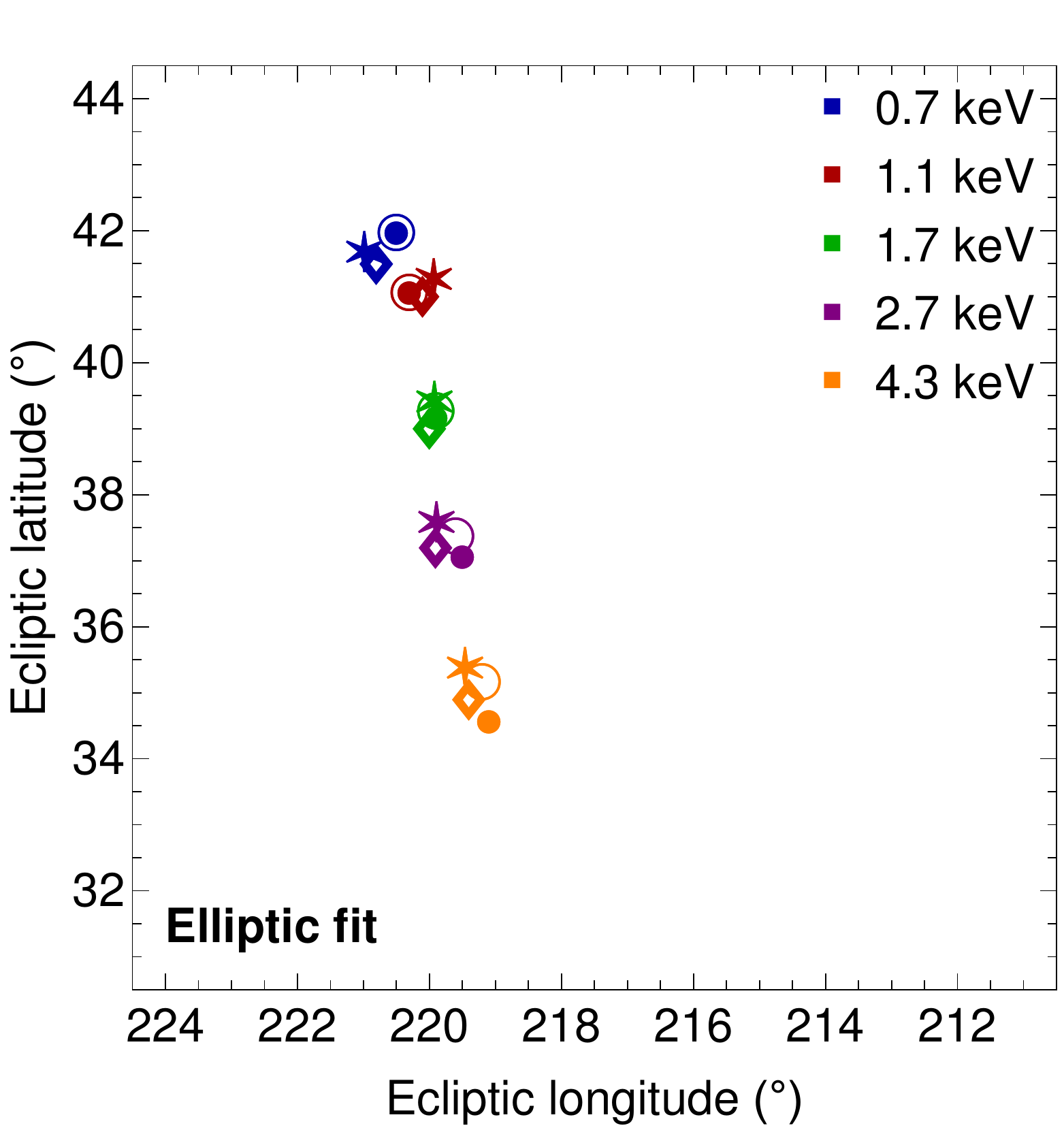}
 \caption{\edit1{Centers of the ellipses fitted to the signal calculated from the model with different sector selections and with or without weighting: stars -- the \citet{funsten_2013} selection with weighting, diamonds -- the \citet{funsten_2013} selection without weighting, open circles -- all sectors included with weighting, filled circles -- all sectors included without weighting.}\label{fig:tests}}
\end{figure}

\edit1{With the presented model, the ribbon peak position can be determined in all azimuthal sectors due to absence of the statistical scatter. With the data, it was not possible, because the ribbon signal is not high enough in some sectors compared to the background. The restriction of the azimuthal sectors, as well as the weighting procedure, can potentially influence the determination of the ribbon center position. We performed three additional fits to quantify this influence. Figure~\ref{fig:tests} shows the positions of the ribbon centers from these fits. Namely, we fit the model with and without the weighting (i.e., $B_i=1$ in Equations~\eqref{chi2c} \&~\eqref{chi2e}), combined with either the selection of sectors made by \citet{funsten_2013} or to all sectors. These modifications shift the ribbon center by at most 1\degr{} and the sequence of the energy channel is in all cases similar. From this test} we conclude that our results and conclusions on the role of the solar wind structure in shaping the position of the ribbon are robust.

The largest discrepancy occurs for the highest energy channel. In the \edit1{model}, the centers for the consecutive energy channels are shifted by the same magnitude for all energy channels, but in the data the highest energy channel is shifted the most, and also the ribbon radius increases accordingly, which is not observed in the \edit1{model}. Another discrepancy is in the fit for the energy 1.7~keV, which in the elliptic case agrees well in all aspects except for the rotation angle (58\degr{} for the \edit1{model} fit and 100\degr{} for the data fit).  Most of the deviations arise due to the statistical dispersion in the data, since the deviations of the ribbon location from the fitted shape ($\sigma_\mathrm{E}$) are 4-fold larger in the data than in the \edit1{model}. The non-vanishing values of $\sigma_\mathrm{E}$ for the \edit1{signal calculated from the model} suggest that the model of a circular or elliptic shape is too simple to adequately describe the ribbon. \edit1{Moreover, we assumed that the geometric factor has maximum along a circle on the sky, while more realistic models of the interstellar magnetic field draping could indicate a more complicated shape. This could be another reason for the discrepancy between the fitted parameters to the model and to the data.} 

\begin{figure*}
\epsscale{1.12}
 \plotone{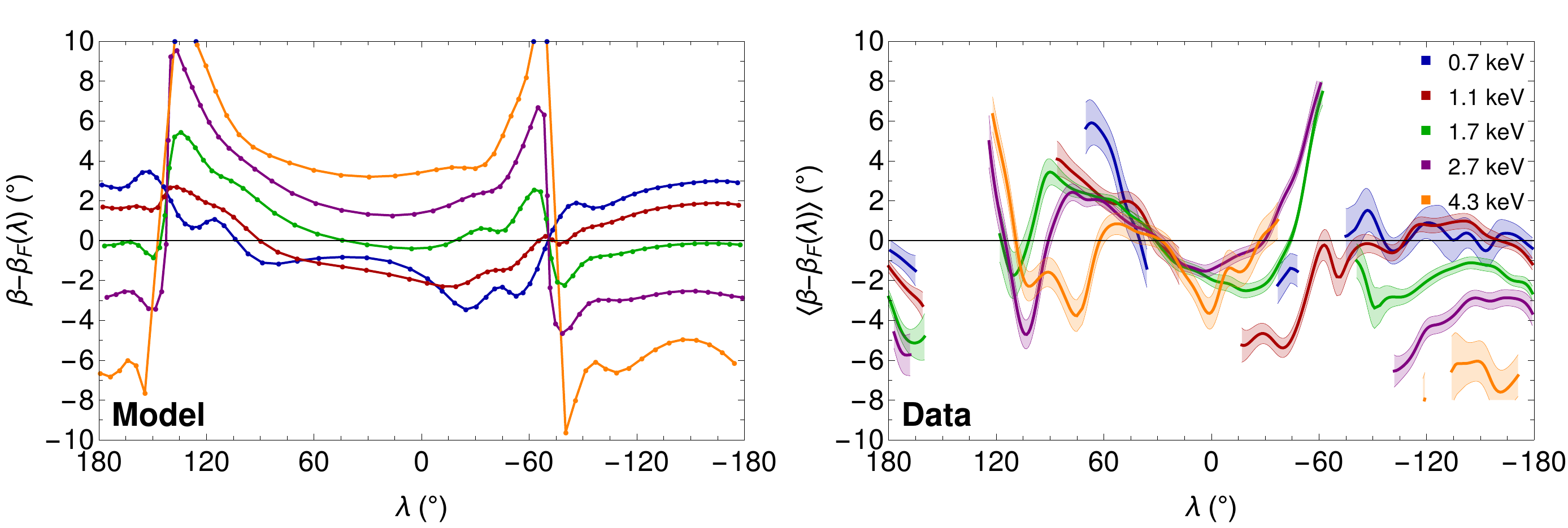} 
 \caption{Comparison of the location of the maximum signal of the ribbon in the ecliptic coordinates for \edit1{the model} (left panel) and the data (right panel) \citep{swaczyna_2016}. The ecliptic latitude defined by the circle centered at $(\lambda_\mathrm{rib},\beta_\mathrm{rib})$ with radius $r_\mathrm{rib}$ is subtracted to highlight the differences between energy channels. Right panel is adopted from Figure~9 in \citet{swaczyna_2016}. Ordinate presents the difference between ecliptic latitude of the ribbon location and the latitude determined from the circle found by \citet{funsten_2013} as average from all energy channels. \label{fig:compareparallax}}
\end{figure*} 

The presented simple model of the secondary ENA mechanism with helio-latitudinal structure of the solar wind reproduces the effect of the energy dependence of the fitted centers of the \emph{IBEX} ribbon very well for most of the \emph{IBEX} energy channels. The discrepancy in the highest energy channel could suggest that for the highest energy at least a portion of the signal may be due to a different mechanism of the ENA generation. An argument in favor of this hypothesis is that the ENA intensities obtained from INCA on board the Cassini spacecraft for energies higher than at \emph{IBEX} reveals a similar feature as the ribbon, called the INCA belt \citep{krimigis_2009,dialynas_2013}, but the center of the belt at $(\lambda_\mathrm{belt},\beta_\mathrm{belt})=(190\degr,15\degr)$ is shifted much farther than the \emph{IBEX} ribbon center. However, the energies of ENAs observed by INCA are well above the energies typical for the solar wind, and thus the belt is not likely explained as the reflectance of the NSW. 

The helio-latitudinal structure of the supersonic solar wind was previously included in several analyses \citep{heerikhuisen_2014,zirnstein_2015,zirnstein_2016a}. However, these analyses did not report any findings concerning the effect of energy-dependent shift of the ribbon center \citep{zirnstein_2015,zirnstein_2016a}, or such an effect was not visible \citep{heerikhuisen_2014}.

As a by-product of the analysis of the ribbon parallax, \citet{swaczyna_2016} obtained deviations of the locations of the maximum signal of the ribbon in the ecliptic coordinates from the positions expected from a circle centered at $(\lambda_\mathrm{rib},\beta_\mathrm{rib})$ with the radius $\phi_\mathrm{rib}$. In Figure~\ref{fig:compareparallax} we compare those results with the deviations obtained in our analysis. In the figure we do not use the fitted shapes, but the actual positions of the maximum signal obtained as the intermediate step in the fitting procedure. The \edit1{model} results cover almost the whole sky, because we can fit the position for any signal level, whereas when fitting the data one needs to adopt a certain threshold value for the signal to noise ratio to find a meaningful fit. 

In the case of a perfectly adequate model the respective lines from the \edit1{model} should fit to the data uncertainty bands, but our model is far too simple to expect a perfect fit. We notice, however, that the energy sequence for the ecliptic longitudes $-120\degr>\lambda>-180\degr$ is the same in the data and in the \edit1{model}. Also the discontinuities for ecliptic longitudes $\sim$75\degr{} and $\sim$--140\degr{} are visible both in the data and in the \edit1{model}. These discontinuities coincide with the intersection of the heliographic equator by the ribbon locations. These results additionally support the connection between the NSW and the \emph{IBEX} Ribbon. 

\section{Summary and conclusions}
\label{sec:conclusions}

In this analysis we extended the analytical model of the secondary ENA emission originally proposed by \citet{mobius_2013}, which we supplemented with the model of the primary ENAs produced in the helio-latitudinally structured supersonic solar wind. The primary ENAs are created by charge-exchange of the solar wind inside the termination shock with the neutral background atoms. The solar wind was modeled using the helio-latitudinal structure from the model by \citet{sokol_2015c}. The distribution of the primary ENAs was built for each CR separately and next averaged over Solar Cycle 23. The obtained signals were subsequently fitted to the circles and ellipses, as was done in the analysis of the \emph{IBEX} data by \citet{funsten_2013}. The fitted parameters, including the centers of the circles and ellipses, were compared between the data and \edit1{model}. 

The ribbon centers for the \emph{IBEX}-Hi energy channels form a monotonic sequence that is well aligned with the \edit1{local heliographic meridian}. The obtained magnitude of this effect is similar to that observed in the data, except for the highest \emph{IBEX}-Hi energy channel, for which the shift between the two highest channels in the data is much larger than for the other pairs of the consecutive channels, what is not observed in the \edit1{model}. This, together with observations of the INCA belt, suggest that in addition to the secondary ENA mechanism that forms the ribbon, a different mechanism may be operating in the vicinity of the heliosphere, responsible for a part of the ENA signal in the highest \emph{IBEX} energy channels and for the INCA belt. 

With the presented model, we reproduced two important features of the ribbon structure: the evolution in the relative magnitude of the signal along the ribbon and the shift of the ribbon center with increasing energy. The first effect was already understood in previous analyses \citep{mccomas_2014b}, but the latter one is explained for the first time. Our findings explain these important features of the ribbon as closely related to each other and strongly support the mechanism of the secondary ENA emission with the interstellar magnetic field lines draped in the outer heliosheath as the most likely mechanism of the ribbon generation. This finding is additionally supported by the distance to the ribbon, determined to be at about 140~AU \citep{swaczyna_2016}. We thus showed that details of the ribbon depend as much on the processes operating in the outer heliosheath as on the details of the solar wind structure and its evolution during the solar cycle. 

\acknowledgements
The authors acknowledge the support by National Science Centre, Poland, grant 2015/19/B/ST9/01328. 

\bibliographystyle{aasjournal}
\bibliography{library}


\end{document}